\documentclass[prl,aps,twocolumn,groupedaddress,floats,showpacs,final,superscriptaddress]{revtex4-2}
\usepackage[latin3]{inputenc}
\usepackage[makeroom]{cancel}
\usepackage{graphicx}
\usepackage{amsmath}
\usepackage{amsfonts}
\usepackage{amssymb}
\usepackage{chngcntr}
\usepackage{color}
\usepackage{cancel}
\usepackage[left=2cm,right=2cm,top=2cm,bottom=2cm]{geometry}

\usepackage{graphicx}
\usepackage{dcolumn}
\usepackage{bm}
\usepackage{simplewick}
\usepackage{array}
\usepackage{appendix}

\RequirePackage[
   hyperindex,colorlinks,bookmarksnumbered,
   plainpages=true,pdfstartview=FitH]{hyperref}
\hypersetup{linkcolor=blue,urlcolor=blue,citecolor=blue}
\usepackage{hyperref}

\definecolor{purple}{rgb}{0.5,0,0.6}

\begin{document}




\title{Proposal for Superconducting Photodiode}




\author{A.~V.~Parafilo}
\email[A.~V.~Parafilo, V.~M.~Kovalev, and I.~G.~Savenko are the corresponding authors: parafilo.sand@gmail.com, vmk111@yandex.ru, ivan.g.savenko@gmail.com]{}
\affiliation{Center for Theoretical Physics of Complex Systems, Institute for Basic Science (IBS), Daejeon 34126, Korea}

\author{M.~Sun}
\affiliation{Faculty of Science, Beijing University of Technology, Beijing, 100124, China}

\author{K.~Sonowal}
\affiliation{Center for Theoretical Physics of Complex Systems, Institute for Basic Science (IBS), Daejeon 34126, Korea}

\author{V.~M.~Kovalev}
\affiliation{Rzhanov Institute of Semiconductor Physics, Siberian Branch\\ of Russian Academy of Science, Novosibirsk 630090, Russia}
\affiliation{Novosibirsk State Technical University, Novosibirsk 630073, Russia}

\author{I.~G.~Savenko}
\affiliation{Department of Physics, Guangdong Technion -- Israel Institute of Technology, 241 Daxue Road, Shantou, Guangdong, China, 515063}
\affiliation{Technion -- Israel Institute of Technology, 32000 Haifa, Israel}

\date{\today}


\begin{abstract}
\textcolor{black}{We propose a concept of a superconducting photodiode -- a device that transforms the energy and `spin' of an external electromagnetic field into the rectified steady-state supercurrent and develop a microscopic theory describing its properties. 
For this, we consider a two-dimensional thin film cooled down below the temperature of superconducting transition with the injected dc supercurrent and exposed to an external electromagnetic field with a frequency smaller than the superconducting gap. 
As a result, we predict the emergence of a photoexcited quasiparticle current, and, as a consequence, oppositely oriented stationary flow of Cooper pairs. 
The strength and direction of this photoinduced supercurrent depend on (i) such material properties as the effective impurity scattering time and the nonequilibrium quasiparticles' energy relaxation time and (ii) such electromagnetic field properties as its frequency and polarization.} 
\end{abstract}

\maketitle

The core of a semiconducting diode is a p-n junction that conducts electric current primarily in one direction thus causing the current rectification effect. 
Diodes represent fundamental building blocks for transistors and conventional integrated electric circuits in general.
A superconducting (SC) diode, instead, is a p-n junction-free valve-effect device, in which it is the current of Cooper pairs (the supercurrent) that flows in one particular direction~\cite{RefScDiods2, RefStrambini}. 
Recent years have brought substantial experimental progress aimed at realizing the current rectification behavior in  superconductors~\cite{JiangHuNature2022,ando,PhysRevLett.128.037001}.
Such devices are providing new opportunities for the design of possible SC circuits. 

Furthermore, a conventional \textit{photodiode} represents a p-n-junction-based semiconductor diode that transforms the energy from an external light source into a rectified electric current.
It is usually tuned to a specific range of frequencies (from IR and THz to the visual light and UV), depending on the material bandgap, and allows for the electromagnetic (EM) field-controlled particle transport.
In this Letter, we develop a theory of nonreciprocal nonlinear photo-response in two-dimensional (2D) superconductors and propose a SC photodiode based on SC thin films or gate-induced superconductors, which have been routinely produced recently by various techniques~\cite{pr10061184, 202006124, Saito_2016, LI2021100504}. 
The realization of a SC photodiode is a direct route towards low-energy small-scale devices with nearly no dissipation involved, fast switch (small inertia), and phase coherence due to the presence of the SC condensate, which, in turn, can provide high sensitivity.

The study of the photoresponse in superconductors exposed to external EM fields is a broad and challenging research topic~\cite{RevModPhys.46.587, 10.1007/978-1-4684-1863-7_9, RevModPhys.77.721, Charnukha_2014, dressel,  PhysRevLett.123.217004}.
On one hand, SC samples weakly interact with (uniform) EM fields; and the photoabsorption is usually suppressed~\cite{Tinkham}. 
On the other hand, it is very tempting to find ways of efficient light-matter interaction in superconductors.
Moreover, the conventional constraints in superconductors are the ambient temperature, external applied magnetic field, and launched electric currents. 
Such characteristics of light as its frequency and polarization can serve as alternative constraints~\cite{PhysRevB.98.064502, PhysRevLett.124.207002}.
\begin{figure*}[t!]
\includegraphics[width=1.8\columnwidth]{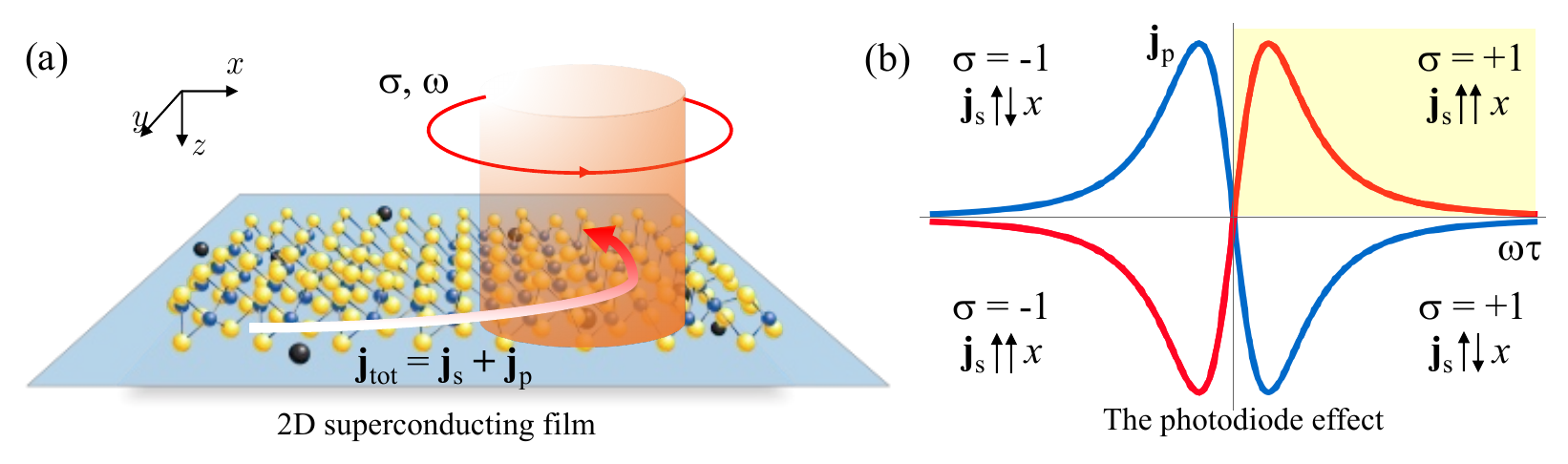} 
\caption{Operating principle of the photodiode. (a) \textcolor{black}{System 
schematic: a 2D superconducting film with a built-in longitudinal (along $x$ axis) dc supercurrent $\textbf{j}_s$ and an external EM field with the circular polarization $\sigma$ and frequency $\omega$  exerted at the normal incidence to the sample plane. Photodiode current flow of Cooper pairs $\textbf{j}_p$ emerges in $y$ direction in order to compensate the current of photo-induced quasiparticles appeared as a consequence of photovoltaic Hall effect.}
(b) Directional pattern of the electric current of the photodiode.}
\label{Fig1}
\end{figure*}

Interested in steady electric current density (time-independent and unidirectional), we will consider the nonlinear (quadratic) in the external EM field amplitude contribution. 
One of the benefits of nonlinear response effects in superconductors, as compared with their linear counterparts, is the possibility of gate-tunability~\cite{RefNonlineHallEff01}, which, by and large, allows for easier access to the p-n-junction-free current rectification~\cite{doi:10.1126/sciadv.aay2497}. 
To demonstrate the sensitivity of the proposed SC photodiode to the frequency and polarization of an external EM field, we study the properties of photoinduced anomalous transverse transport of Cooper pairs, when an external circularly-polarized THz field is exerted on the sample.

An attempt to make a superconductor interact with a uniform external EM field is only seemingly easy.
One of the problems is that in clean single-band~\cite{PhysRevB.95.014506,NagaosaOptRe2021, PhysRevB.106.214526,PhysRevB.106.094505,PhysRevB.108.224516} conventional superconductors, the photoabsorption is suppressed for symmetry reasons~\cite{PhysRev.108.1175, Mahan}.
However, the presence of impurities can lift this problem~\cite{PhysRev.111.412, PhysRev.156.470, PhysRev.156.487,ZIMMERMANN199199, PhysRev.165.588}.
Furthermore, a stationary transverse photocurrent represents a second-order response to the external EM field, 
determined by a third-rank tensor $\eta_{\alpha\beta\gamma}$: $j_\alpha=\eta_{\alpha\beta\gamma}E_\beta E^*_\gamma$, where $E_\beta$ are the components of electric field amplitude
~\cite{[{In the expression $j_\alpha=\eta_{\alpha\beta\gamma}E_\beta E^*_\gamma$ a standart Einstein convention is used: the summation over indexes $\beta$ and $\gamma$ is assumed. A third-rank tensor here is nonzero only in the media lacking the space inversion. In our consideration, the built-in supercurrent determined by the supermomentum ${\bf p}_s$ destroys the space inversion, thus, a tensor $\eta_{\alpha\beta\gamma}[{\bf p}_s]$ directly depends on ${\bf p}_s$. For small values of supercurrents, one can expand $\eta_{\alpha\beta\gamma}[{\bf p}_s]$ up to linear term with respect to ${\bf p}_s$, and for isotropic superconductors and circularly-polarized illumintaion one finds Eq.~\eqref{current} for the current}]CommentTensor}.
This tensor is only nonzero in the media lacking the inversion symmetry. 
For isotropic superconductors, a possible way to destroy the inversion symmetry is to launch a built-in supercurrent~\cite{PhysRevB.106.L220504, PhysRevLett.122.257001}.



\textit{Phenomenological description.---}
Let us consider an isotropic 2D layer below the SC critical temperature $T_c$ but at a finite temperature $T$ and with a given (source) stationary current of Cooper pairs $\textbf{j}_s$ in the $x$-direction [Fig.~\ref{Fig1}(a)]. 
Furthermore, the sample is illuminated by an external EM field, described by the vector-potential ${\bf \mathcal{A}}(t)={\bf \mathcal{A}}\exp{(-i\omega t)}+{\bf \mathcal{A}}^*\exp{(i\omega t)}$ at a normal incidence~\cite{[{The developed theory is gauge invariant in the xy plane. In derivations, we used the gauge $\varphi = 0$ for the EM field with a normal incidence to the 2D plane, and considered the second-order response, which is transverse to the direction of EM wave propagation. In the case of an oblique incidence of the external EM field, the gauge invariance requires taking into account the collective excitations of the order parameter~\cite{Arseev_2006}. Moreover, an effect of photon drag with momentum transfer from EM wave to the condensate should be accounted for, see, e.g.,~\cite{PhysRevB.106.094505,PhysRevLett.132.096001}}]DRAG}. 
Since there is a finite amount of thermally excited quasiparticles above the SC gap $\Delta$ at $T\neq0$, we can restrict ourselves to considering $\omega\ll \Delta/\hbar$.
The EM field is circularly polarized, $\mathcal{A}=\mathcal{A}_0(1, i\sigma)$, where $\sigma=\pm 1$ stands for the left/right circular polarization.
As a result of the light absorption by the sample, a photoinduced electric current of electrons above the SC gap emerges, the stationary part of which reads
\begin{gather}
\label{current}
{\bf j}=
ic_\omega[{\bf p}_s\times[{\bf \mathcal{A}}\times{\bf \mathcal{A}}^*]],
\end{gather}
where ${\bf p}_s=(p_s,0)$ is the Cooper pairs' momentum.
Phenomenological expression~\eqref{current} describes the stationary photoinduced electric current in the system due to the emergence of a correction to the electron distribution function,  $\delta f(\mathbf{v},\mathbf{E},\mathbf{p}_s)\propto({\bf v}\cdot{\bf E})({\bf v}\cdot{\bf E}^*)({\bf v}\cdot{\bf p}_s)$, which refers to the phenomenon called \textit{the anisotropic alignment of photoelectrons}~\cite{PhysRevB.104.085306}.
It provides the accumulation of positive and negative charges at the transverse boundaries of the sample.

Since the electric field nearly cannot penetrate the SC sample, the current of photoinduced quasiparticles~\eqref{current} should be accompanied by an induced condensate flow so that the electric field is compensated, 
following the general properties of superconductors~\cite{Shmidt, PhysRevLett.125.097004, PhysRevB.18.5116, PhysRevB.21.1842,PhysRevB.65.064531}. 
As a result, there emerges a flow of a frequency- and polarization-controlled unidirectional electric current of Cooper pairs $\mathbf{j}_\textrm{p}$. 
Indeed, the photoinduced quasiparticle current \eqref{current} as a response to the circularly-polarized EM field, must vanish both at zero frequencies, $\omega\rightarrow0$, and large frequencies, $\omega\rightarrow\infty$. 
It also changes its sign if the built-in supercurrent changes its direction to the opposite one, ${\bf j}_s\rightarrow-{\bf j}_s$. Taking into account that the frequency reversing $\omega\rightarrow-\omega$ corresponds to the change of external EM field polarization, $(\omega\leftrightarrow-\omega)\Leftrightarrow(\sigma\leftrightarrow-\sigma)$, we come up with the  polarization/supercurrent pattern of SC photodiode operation, presented in Fig.~\ref{Fig1}(b). 
This figure demonstrate the sensitivity of the rectified condensate photocurrent to the direction of built-in supercurrent and light polarization, constituting the SC photodiode effect. 
Below, we develop the microscopic description of the photodiode to support our phenomenological considerations and explore its properties, analyse the pattern $(\sigma,{\bf p}_s,\omega\tau)$, and investigate the origin of one of the key entities: the relaxation time $\tau$.
\begin{figure*}[t]
\includegraphics[width=1.8\columnwidth]{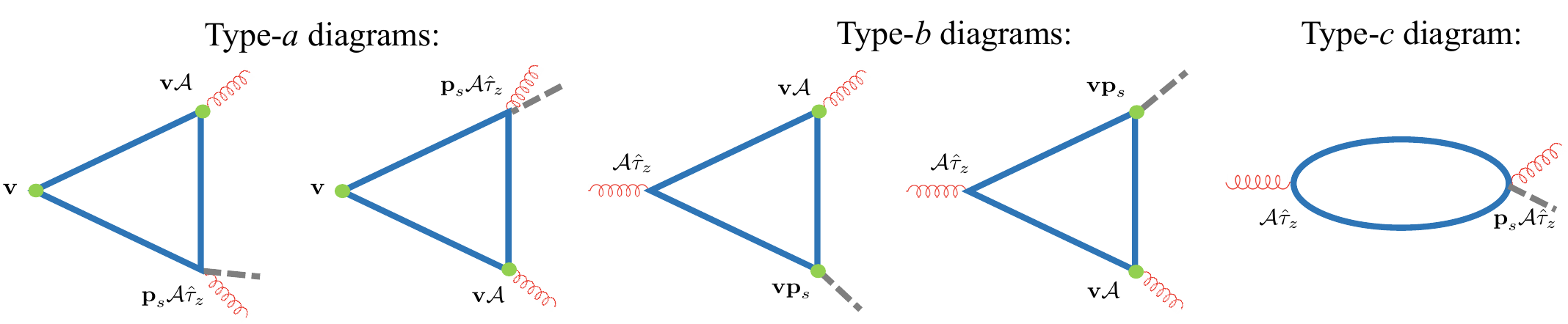} 
\caption{The highest-contribution Feynman diagrams describing the photoinduced quasiparticle current. 
Blue lines are the Green's functions of quasiparticles, red springs describe the external EM field $\mathcal{A}$; green dots stand for the quasiparticle velocity vertices $\textbf{v}$, and dashed lines show the supercurrent momentum ${\bf p}_s$.}
\label{Fig2}
\end{figure*}
%
%
%
%



\textit{Microscopic description.---}
The Hamiltonian of a single-band isotropic 2D $s$-wave superconductor exposed to an external EM field reads
($\hbar=k_B=c=1$)
\begin{gather}\label{Ham}
\hat{H}=
\left(\begin{array}{cc}
\xi\left({\bf p}-{\bf p}_s-e\mathcal{A}(t)\right) & \Delta \\
\Delta  & -\xi\left(-{\bf p}-{\bf p}_s-e\mathcal{A}(t)\right)
\end{array}\right),
\end{gather}
where $\xi(\textbf{p})\equiv\xi_\textbf{p}=\textbf{p}^2/2m-E_F$ is the electron kinetic energy, $E_F$ is the Fermi energy, and $\Delta$ is a SC gap, which we assume to be real-valued. 
The electric current density is determined as the trace of the current operator with a ``lesser'' component of the Green's function,
\begin{eqnarray}\label{Current}
{\bf j}(t)=-i\,\textmd{Tr}\left\{\hat{{\bf j}}\,\hat{\mathcal{G}}^{<}(t,t)\right\}.
\end{eqnarray}
In Eq.~\eqref{Current}, the operator of the current represents the variation of the Hamiltonian~(\ref{Ham}) over the vector-potential,
\begin{eqnarray}
\hat{\textbf{j}}=-\frac{\delta \hat{H}}{\delta \mathcal{A}}=e\textbf{v}-\frac{e\textbf{p}_s}{m}\hat{\tau}_z-\frac{e^2\mathcal{A}}{m}\hat{\tau}_z,
\end{eqnarray}
and $\hat{\mathcal{G}}^<(t,t')$ is the lesser component of the Keldysh Green's function $\hat{\mathcal{G}}(t,t')$ determined by the equation $(i\partial_t-\hat{H})\hat{\mathcal{G}}(t, t')=\delta(t-t')$.

Expanding $\hat{\mathcal{G}}(t,t')$ up to the first order with respect to $\textbf{p}_s$ and up to the second order with respect to $\mathcal{A}(t)$ yields the total current density. 
The main contribution to the transverse component of the current density (for the case of circularly polarized light) can be described by five diagrams, split into three types (Fig.~\ref{Fig2}).
The resulting components of the current density read
\begin{eqnarray}\label{curr1}
\textbf{j}^{(a,b)}&=&ie^3\sum_{
\textbf{p}}\int dt_1\int dt_2 
\textrm{Tr}\left\{\hat{\gamma}^{(a,b)}\hat{g}(t,t_1)\right.\\
\nonumber
&&~~~~~~~~~~~~~~~~~\left.\times\hat{\gamma}^{(a,b)}_{\pm}\hat{g}(t_1,t_2)\hat{\gamma}^{(a,b)}_{\mp}\hat{g}(t_2,t)\right\}^<,\\
\textbf{j}^{(c)}&=&\frac{ie^3}{m}\sum_{
\textbf{p}}\int dt_1\textrm{Tr}\left\{\hat{\gamma}^{(c)}\hat{g}(t,t_1)\hat{\gamma}^{(c)}_{-}\hat{g}(t_1,t)\right\}^<,
\label{curr2}
\end{eqnarray}
where $\hat{g}(t,t')$ is a bare Keldysh Green's function, which represents a solution of the equation $(i\partial_t-\hat{H})\hat{g}(t, t')=\delta(t-t')$ with the Hamiltonian~\eqref{Ham}, in which $\textbf{p}_s=0$ and $\mathcal{A}_0=0$; $\gamma^{(a,b,c)}_{\,\,,+,-}$ are generalised vertices: $\hat{\gamma}^{(a)}$$=$$\textbf{v}$, $\hat{\gamma}^{(b,c)}$$=$$\mathcal{A}\hat{\tau}_z$, $\hat{\gamma}^{(a,b)}_{+}$$=$$(\textbf{v}\cdot\mathcal{A})$, $\hat{\gamma}^{(a,c)}_{-}$$=$$(\textbf{v}_s\cdot\mathcal{A})\hat{\tau}_z$, $\hat{\gamma}^{(b)}_{-}$$=$$(\textbf{v}\cdot\textbf{v}_s)$, with $\textbf{v}_s$$=$$\textbf{p}_s/m$.
Furthermore, in Eqs.~\eqref{curr1} and~\eqref{curr2} let us shift to the energy-momentum representation and consider the first stationary photo-induced correction to the current density. 
Direct calculation of the current density using the bare Green's function $\hat{g}_{\epsilon}=(\epsilon-\Delta \hat{\tau}_x-\xi_{\textbf{p}}\hat{\tau}_z)^{-1}$ gives a vanishing contribution, as expected: 
The selection rules forbid optical absorption in the superconductor due to preserving the Galilean invariance even in the case of broken spatial symmetry ($\textbf{p}_s\neq 0$)~\cite{PhysRevB.95.014506}.  

The Galilean invariance can be broken by an account of either i) non-parabolicity of the electronic dispersion~\cite{PhysRevB.95.014506}, or ii) multi-bands~\cite{NagaosaOptRe2021,PhysRevB.106.214526}, or iii) impurities~\cite{PhysRevB.106.L220504,PhysRevB.108.L180509}.
We will use the latter mechanism (the scattering on non-magnetic impurities). 
Solving the Dyson equation and averaging over the disorder potential~\cite{Abrikosov:107441}, we find the retarded, advanced, and lesser Green's function of electrons in 2D superconductor:
\begin{eqnarray}\label{RetGr}
\hat{g}^{R}_{\epsilon}&=&\frac{\epsilon \eta_{\epsilon}+\xi_{\textbf{p}} \hat{\tau}_z+\Delta \eta_{\epsilon}\hat{\tau}_x}{\left(\epsilon-\epsilon_{\textbf{p}}+ \frac{i}{2\tau_{\textbf{p}}}\right)
\left(\epsilon+\epsilon_{\textbf{p}}+\frac{i}{2\tau_{\textbf{p}}}\right)}, \\\nonumber
\hat{g}^A_{\epsilon}&=&\left(\hat{g}^{R}_{\epsilon}\right)^{\ast},~~~~~~~~~~
\hat{g}_{\epsilon}^<=-f_{\epsilon}\left(\hat{g}^R_{\epsilon}-\hat{g}^{A}_{\epsilon}\right).
\end{eqnarray}
Here, $\epsilon_{\textbf{p}}=\sqrt{\xi_{\textbf{p}}^2+\Delta^2}$ is the quasiparticles dispersion, $\tau_{\textbf{p}}$ is their effective lifetime, $f_{\epsilon}=[\exp(\epsilon/T)+1]^{-1}$ is the Fermi-Dirac distribution function,
\begin{eqnarray}\label{eta}
&&\eta_{\epsilon}=1+\frac{i}{2\tau_i}\frac{\textrm{sign}[\epsilon]}{\sqrt{\epsilon^2-\Delta^2}}
\end{eqnarray}
is a renormalization factor, which accounts for the disorder with $\tau_i=(2\pi \nu_0 n u_0^2)^{-1}$ the characteristic time of electron scattering on impurities  with $\nu_0$ is the density of states of electron gas in non-SC state, $n$ is the density of impurities, $u_0$ is the impurity electrostatic potential~\cite{Abrikosov:107441}. 

However, the elastic scattering (characterized by $\tau_i$) is not the only mechanism, and, in general, various relaxation processes may play role in a SC sample exposed to an EM field. Which particular relaxation time makes a dominant contribution depends on the frequency of the EM field, the properties of the SC sample, and how the frequency is related to the SC gap~\cite{PhysRevB.101.134508, SMITH2020168105,Ovchinnlsaakyan}. At small frequencies $\omega\ll\Delta$ and the temperatures in the vicinity of the SC critical temperature $T_c$, the optical conductivity 
depends on the inelastic relaxation time $\tau_E$, which is, in turn, determined by the energy relaxation of quasiparticles in 2D films~\cite{Ovchinnlsaakyan}.
Altogether, the relaxation processes are described by the effective inverse lifetime,
\begin{gather}
\label{totaltime}
\frac{1}{\tau_{\textbf{p}}}=
\frac{1}{\tau_E}
+
\frac{1}{\tau_i}\frac{|\xi_{\textbf{p}}|}{\epsilon_{\textbf{p}}}.
\end{gather}
%
%
%
%
\begin{figure}[t!]
\includegraphics[width=0.99\columnwidth]{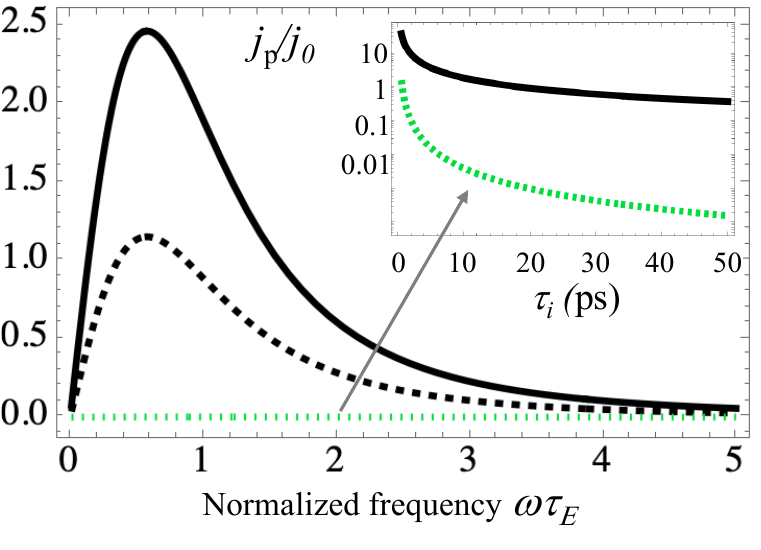}
\caption{Spectrum of the photodiode current: photoinduced current density as a function of frequency of the external EM field.
Black solid and dashed curves correspond to the contributions of type-a and type-b diagrams [see Eq.~(\ref{expr1})], for $T=5$~K and $T=3.5$~K, respectively; green dotted curve illustrates the contribution of type-c diagram at $T=5$~K [Eq.~(\ref{expr2})].
We used $\Delta_0=0.8$~meV ($T_c\approx 5.67$~K), $\tau_i=10$~ps, and $\tau_E=100$~ps.
Inset shows the maximum current density for type-a and type-b diagrams (black) and type-c diagram (green) as functions of $\tau_i$ for  $\tau_E=100$~ps and $T=5$~K.
}
\label{Fig3}
\end{figure}
%
%
%

Substituting $\tau_\mathbf{p}$ in equations above and performing calculations (see the details in Supplemental Material~\cite{[{See Supplemental Material at [URL], which gives the details of the analysis of all the relevant Feynman diagrams}]SMBG}), we find 
%
\begin{eqnarray}\label{expr1}
j_y^{(a)}+j_y^{(b)}= \frac{4j_0\omega\tau_E}{\left[1+(\omega\tau_E)^2\right]^2}\frac{\tau_E}{\tau_i}\left(\frac{\Delta}{2T}\right)^2\mathcal{I}_1\left(\frac{\Delta}{2T}\right),\\ 
j_y^{(c)}=\frac{j_0}{2} \frac{\omega\tau_E}{1+(\omega\tau_E)^2}\frac{1}{(2T\tau_i)^{2}}\left(\frac{\Delta}{2T}\right)^2\mathcal{I}_2\left(\frac{\Delta}{2T}\right),\label{expr2}
\end{eqnarray}
where $j_0=\sigma (e^3p_s\mathcal{A}_0^2/m\pi)$,
and the dimensionless integrals read
\begin{eqnarray}
&&\mathcal{I}_1(y)=\int_{y}^{\infty} \frac{dx}{x^2\cosh^2(x)},\\
&&\mathcal{I}_2(y)=\int_{y}^{\infty}\frac{dx}{x^3\cosh^2(x)} \frac{1}{\sqrt{x^2-y^2}}.
\end{eqnarray}
Eqs.~\eqref{expr1} and~\eqref{expr2} also account for the temperature dependence of the SC gap, $\Delta(T)= \Delta_0\tanh\left(1.76\sqrt{T_c/T-1}\right)$ with $\Delta_0$$\equiv$$\Delta(T$$=$$0)$ and $T_c=\Delta_0(\exp[0.577]/\pi)$. 
The emergent photodiode supercurrent reads as $j_p=-\left(j_y^{(a)}+j_y^{(b)}+j_y^{(c)}\right)$.




\textit{Results and discussion.---} 
Formulas~\eqref{expr1} and~\eqref{expr2} describe the main contributions to the photodiode current density and constitute the main result of this Letter. 
Let us note that both of these contributions vanish as $\tau_i\rightarrow\infty$, as expected.
Indeed, the photoabsorption and hence, the photoinduced quasiparticle transport should be absent in a clean sample, as required by the Galilean invariance~\cite{[{More specifically, in Eq.~(\ref{curr2}) there also appears a non-dissipative term, which does not depend on $\tau_i$. This is a well-known problem that occurs since the BCS theory is non-gauge invariant. The gauge invariance can be restored by accounting for the vertex correction associated with the BCS  interactions~\cite{PhysRev.117.648, PhysRevB.95.014506}. Here, we skip this consideration by simply presenting Eq.~(\ref{expr2}) as $j^{(c)}_y=j^{(c)}_y(\tau_i\neq \infty)-j^{(c)}_y(\tau_i\rightarrow \infty)$}]GAUGE}.
\textcolor{black}{We also note, that formulas~(\ref{expr1}) and~(\ref{expr2}) are universal: they can be applied to a wide range of SC thin films or two-dimensional transition metal dichalcogenides-based superconductors~\cite{pr10061184, 202006124, Saito_2016,LI2021100504}. Evidently, the strength of the photodiode current is determined by the ratio of the inelastic and impurity scattering relaxation times and the value of injected dc supercurrent $\textbf{j}_s$.} 

Let us estimate the magnitude of the photodiode current and compare it with the magnitude of the built-in (longitudinal) supercurrent. 
Writing the Cooper pair momentum as $p_s/m=j_s/(en_s)$, where $n_s(T)$ is the density of particles in the condensate (here,  $n_s(T)=n(1-T/T_c)=7.1\cdot 10^{12}$~cm$^{-2}$ at $T=5$~K for superconductor with $T_c=5.67$~K and $n=0.6\cdot 10^{14}$ cm$^{-2}$), and taking $\omega\approx20$~GHz, $\tau_E=10^{-10}$~s (for typical energy relaxation time see, e.g.,~\cite{PhysRevB.102.054501}), $\tau_i=10^{-11}$~s, and for the EM field intensity $I=1$~W/cm$^2$, we find $j_p/j_s\approx 0.12$. 
Thus, the transverse current represents a considerable correction to the built-in current. \textcolor{black}{For the estimations, we took a typical intensity of EM field from Ref.~\cite{PhysRevLett.125.097004}. Here, we considered a MoS$_2$-based gate-induced superconductor as a testbed to observe the superconducting photodiode effect.
Other typical parameters, such as $T_c$ and typical particle densities (in a MoS$_2$-based superconductor), were taken from Ref.~\cite{doi:10.1126/science.aab2277}. 
It should be noted, that the frequency of the EM field is restricted by the approximations we used in the paper: the frequence of the wave obeys $\omega\ll \Delta$; the impurity relaxation time is conditioned by $\Delta\tau_i\gg 1$ (this restriction is associated with the applicability of the diagram technique).}

\textcolor{black}{Similarly, we can make estimations for other 2D SC materials, for instance, the ones based on NbN thin films or other transition metal dichalcogenide-based systems. 
Other typical examples are summarized in Table below:
} 
\begin{center}
\begin{tabular}{ |c|c|c| } 
 \hline
 \textbf{Material} & \textbf{Critical temperature} & \textbf{SC gap} \\ 
 NbN thin film & 8-15~K & 1.2-2.2 meV \\ 
 MoS$_2$ & 2-10 K & 0.3-1.5 meV\\
 MoSe$_2$ & 7-10 K & 1-1.5 meV \\
 NbSe$_2$ & 7 K & 1 meV \\
 TaSe$_2$ & 0.2-0.3 K & 30-45 $\mu$eV \\
 MoTe$_2$ & 0.1-0.2 K & 15-30 $\mu$eV \\
 \hline
\end{tabular}
\end{center}
\textcolor{black}{For all these materials, typical electron densities are $10^{13}-10^{14}$~cm$^{-2}$.
The density can vary depending on such factors as the method of fabrication, doping levels, and the growing conditions. Besides, the values of critical temperatures and SC gap are of the same order. Thus, we conclude that all mentioned in the Table materials could be efficiently used as a base for SC photodiode.}

It should be noted, that the predicted photodiode effect (i.e. the emergence of a photoexcited quasiparticle current $\propto \Delta^2(T)$, and, as a consequence, oppositely oriented supercurrent in the direction transverse to the built-in supercurrent) can only occur in the SC state, namely, at $T<T_c$. 
Why is that significant?
Let us elaborate on that.
The normal-state (non-SC) counterpart of the SC photodiode effect is a Photovoltaic Hall effect~\cite{[{The photovoltaic Hall effect is an effect consisting in the appearance of finite transverse conductivity under the influence of circularly or linearly polarized radiation in a 2DEG in the direction transverse to the in-plane electric field}]PHE}. 
This effect takes place if the electron relaxation time depends on energy.  
For that, usually the presence of Coulomb scatterers (long-range impurities) in the sample is required. 
In the SC sample, however, even short-range scatterers are sufficient due to the appearance of the SC energy-dependent density of states in front of the impurity scattering time, $\tau_i \left(\epsilon_{\textbf{p}}/\sqrt{\epsilon_{\textbf{p}}^2-\Delta^2}\right)$, see Eq.~(\ref{RetGr}) and the second term in r.h.s. of Eq.~(\ref{totaltime}). 
Hence, effect vanishes at $T>T_c$ when $\tau_i \epsilon_{\textbf{p}}/\xi_{\textbf{p}}\rightarrow \tau_i$, since $\Delta(T)=0$.
\begin{figure}[t!]
\includegraphics[width=0.99\columnwidth]{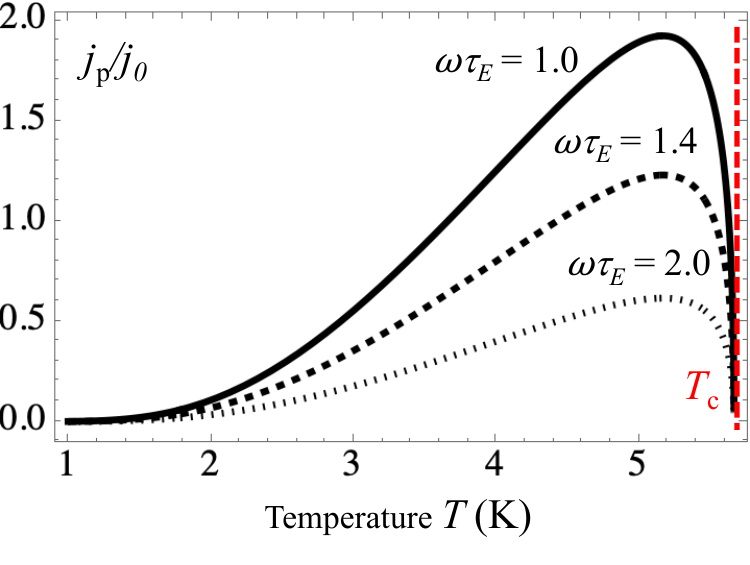}
\caption{Temperature dependence of the photodiode current density for different frequencies of the external EM field.
The parameters were taken the same as in the main panel of Fig.~\ref{Fig3}. \textcolor{black}{Red dashed line indicates the SC critical temperature at $T_c\approx 5.67$ for $\Delta_0=0.8$ meV.}
}
\label{Fig4}
\end{figure}
%
%
%


Figure~\ref{Fig3} shows the spectrum of normalized by $j_0$ electric current density of Cooper pairs for $\Delta_0=10\,\textrm{K} (0.8 \,\textrm{meV}) $.
The largest values for the current correspond to the frequencies $\omega\approx 0.577 \tau_E^{-1}$. The main contribution to the transverse current density originates from the type (a) and (b) diagrams since they are determined by the ratio $\tau_E/\tau_i$. At the same time, the contribution of the type (c) diagram depicted in Fig.~\ref{Fig2} is proportional to $(\tau_i T_c)^{-2}$ (see Eq.~(\ref{expr2})) and hence it is small since $\Delta\tau_i \gg 1$ for the given set of parameters, as compared to other diagrams. However,
the situation is different for the dirtier SC samples when $\Delta\tau_i\lesssim 1$, see inset in Fig.~\ref{Fig3}: The contribution from diagram (c) increases with the decrease of influence of elastic scattering and might become dominant.

Figure~\ref{Fig4} shows the temperature dependence of total photodiode current. 
For the temperatures approaching absolute zero, the current vanishes, as expected, since thermally-excited quasiparticles disappear.
The current also vanishes when approaching the temperature of the SC transition, since the Cooper pairs disappear at $\Delta(T)\rightarrow 0$. 
We also note the increase of the maximum value of the current with the decrease of frequency and the presence of the optimum.


\textcolor{black}{
One of the possible experiment to measure the proposed effect involves using a SC loop that encloses the transverse boundaries of the sample (in y-direction). Due to the photoinduced phase difference associated with the supercurrent flow in $y$-direction, there emerges a correction to the flux through the loop, $\Phi=n\Phi_0+\delta\Phi$, $\delta\Phi\propto j_p$, where  $\Phi_0$ is the flux quantum and $n$ is an integer. The methods to measure such a flux are well developed; they have been  used, e.g., to study the thermoelectric effect-induced condensate phase difference~\cite{PhysRevB.21.1842}.
They can also be employed for the detection of acoustoelectric phenomena in superconductors~\cite{PhysRevB.18.5116}. An alternative is to attach a normal metal to the sample to measure the photoinduced quasiparticle current at the normal metal-insulator-superconductor junction~\cite{PhysRevB.25.4515}. This method has also proved efficient for the measurement of the thermoelectric effect-induced condensate phase difference~\cite{PhysRevLett.39.660}. Another alternative is the optical methods~\cite{PhysRevLett.122.257001, PhysRevLett.124.087701}, namely, the Kerr effect, based on the measurement of the change in the polarization of the reflected or transmitted EM field~\cite{PhysRevB.92.100506, PhysRevB.108.104506}.
}




\textit{Conclusions.---} 
We developed a nonequilibrium theory of nonlinear photoresponse in a two-dimensional superconductor, exposed to a circularly polarized light with a terahertz frequency, which is smaller than the superconducting gap.
The resulting rectified photoinduced electric current of Cooper pairs represents the second-order correction with respect to the vector potential of light. 
This current is determined by the energy relaxation in the system and the presence of impurities, and its magnitude and direction can be controlled by the frequency and polarization of light. 
Thus, we proposed a photodiode based on two-dimensional superconductors.

\textit{Acknowledgements.} 
The authors are grateful to Prof. Boris Altshuler for useful discussions.
We were supported by the Institute for Basic Science in Korea (Project No.~IBS-R024-D1), 
the Foundation for the Advancement of Theoretical Physics and Mathematics ``BASIS''. 
M. S. is also supported by the R\&D Program of Beijing Municipal Education Commission ( KM202410005011).

\bibliography{biblio}
\bibliographystyle{apsrev4-2}


\end{document}